\RequirePackage{fix-cm}
\documentclass[smallextended]{svjour3}       % onecolumn (second format)
\smartqed  % flush right qed marks, e.g. at end of proof
\usepackage{graphicx}
\begin{document}

\title{Gravity, antimatter and the Dirac-Milne universe%\thanks{Grants or other notes
}
%\subtitle{}


\author{Gabriel Chardin         \and
        Giovanni Manfredi %etc.
}


\institute{G. Chardin \at
              CNRS, 3 rue Michel-Ange, F-75016 Paris, France \\
              \email{gabriel.chardin@cnrs-dir.fr}           %  \\
           \and
           G. Manfredi \at
Universit\'e de Strasbourg, CNRS, IPCMS UMR 7504, F-67000 Strasbourg, France.\\
\email{giovanni.manfredi@ipcms.unistra.fr}
}

\date{Received: date / Accepted: date}

\maketitle

\begin{abstract}
We review the main arguments against antigravity, a different acceleration of antimatter relative to matter in a gravitational field, discussing and challenging Morrison's, Good's and Schiff's arguments. Following Price, we show that, very surprisingly, the usual expression of the Equivalence Principle is violated by General Relativity when particles of negative mass are supposed to exist, which may provide a fundamental explanation of MOND phenomenology, obviating the need for Dark Matter.

Motivated by the observation of repulsive gravity under the form of Dark Energy, and by the fact that our universe looks very similar to a coasting (neither decelerating nor accelerating) universe, we study the Dirac-Milne cosmology, a symmetric matter-antimatter cosmology where antiparticles have the same gravitational properties as holes in a semiconductor. Noting the similarities with our universe (age, SN1a luminosity distance, nucleosynthesis, CMB angular scale), we focus our attention on structure formation mechanisms, finding strong similarities with our universe.

Additional tests of the Dirac-Milne cosmology are briefly reviewed, and we finally note that a crucial test of the Dirac-Milne cosmology will be soon realized at CERN next to the ELENA antiproton decelerator, possibly as early as fall 2018, with the AEgIS, ALPHA-g and Gbar antihydrogen gravity experiments.

\end{abstract}

\section{Introduction}
\label{sec:intro}
The vast majority of theoretical physicists believe that, if a difference in acceleration between matter and antimatter exists, it can only be extremely small. Few consider possible an antigravity where antihydrogen would ``fall up", as the CERN presents the three current experiments testing the equivalence principle by the ELENA antiproton decelerator \cite{CERN_courrier}.

Why then these three experiments, AEgIS \cite{Kellerbauer}, ALPHA-g \cite{Bertsche} and Gbar \cite{Indelicato}, aim only at a precision of the order of one percent, at least in a first stage, while at the same time, the BASE experiment \cite{BASE_2017} claims to impose constraints on any anomalous gravity for antimatter at the sub-ppm level? Could it be that antigravity, in the sense of antimatter ``falling up", is actually a prediction, which seems at first antinomic, of general relativity?

In a first part of this work, we briefly review the impossibility arguments against antigravity, focusing on Schiff's \cite{Schiff}, Morrison's \cite{Morrison} and Good's \cite{Good} arguments, showing why they are probably ineffective. We then discuss, also rather briefly, the so-called Klein paradox, or vacuum polarization, which provides some elements of answer concerning the impossibility of negative energy states and negative mass. More fundamentally, we describe the argument by Price \cite{Price} showing that general relativity violates maximally the usual expression of the Equivalence Principle as soon as the existence of negative mass, possibly as virtual constituents of the quantum vacuum, is allowed.

This will lead us to the Dirac-Milne universe \cite{Benoit-Levy_Chardin}, and a possible explanation of the repulsive gravity that we observe in cosmology, called Dark Energy for lack of better comprehension. This matter-antimatter universe is impressively concordant, and has also, a fact that is often not realized, a simple physical analog with the electron-hole system in a semiconductor.

Next, we discuss the mechanism of structure formation in the Dirac-Milne universe, radically different from that of the Lambda-CDM universe, and show that, without any free parameter, it reproduces several of the features observed in large surveys such as SDSS \cite{SDSS}.

In a final part, we discuss the additional experiments and studies that can be realized in the near future to test the Dirac-Milne cosmology.

\section{Impossibility arguments}
\label{sec:imposs}
Over the years, several impossibility arguments have been raised against antigravity. A rather thorough discussion of the main impossibility arguments can be found in the review by Nieto and Goldman \cite{Nieto_Goldman}, dating back to 1991 but still mostly valid today. Truly enough, as soon as we express general relativity as a metric theory, with a single metric, it is difficult to see how gravity could distinguish matter from antimatter since according to the very formalisme of a single metric, all particles must follow the same trajectory. Still we will see that General Relativity does predict gross violations of the Equivalence Principle as soon as negative mass components are allowed.
Also, as was noted by physicists in solid-state physics \cite{Tsidil} and in structure formation \cite{Dubinski_Piran,Piran}, there could be other expressions of the Equivalence Principle respecting the spirit of General Relativity but violating maximally its usual expression.
Coming back to the impossibility arguments, we can summarize them in three classes, that we might call the Morrison argument \cite{Morrison}, the Schiff argument \cite{Schiff}, and the Good argument \cite{Good}.

\subsection{Morrison's argument}
As early as 1958, Morrison, in a celebrated paper associated with his Richtmyer memorial lecture \cite{Morrison}, studied the consequences of antigravity in a gedanken experiment that can be summarized in Fig. 1. Basically, the argument states that, if we accept antigravity, energy is not conserved and/or the vacuum becomes unstable. The question(s) that Morrison did not ask was : ``Unstable, by how much, and what is the characteristic timescale?"

Noting that in some other situations, the vacuum of gravitational structures such as black holes is unstable, since black holes evaporate, one of us tried to estimate this instability at the beginning of the 1990s \cite{Chardin_Rax}. Remarkably, the instability that we can expect from antigravity is the same as the Hawking evaporation of black holes, and leads to a temperature of radiation of:
$$k_B T \approx \hbar g/2\pi c$$
where $g$ is the usual surface gravity.

\begin{figure}\sidecaption
\resizebox{0.6\hsize}{!}{\includegraphics*{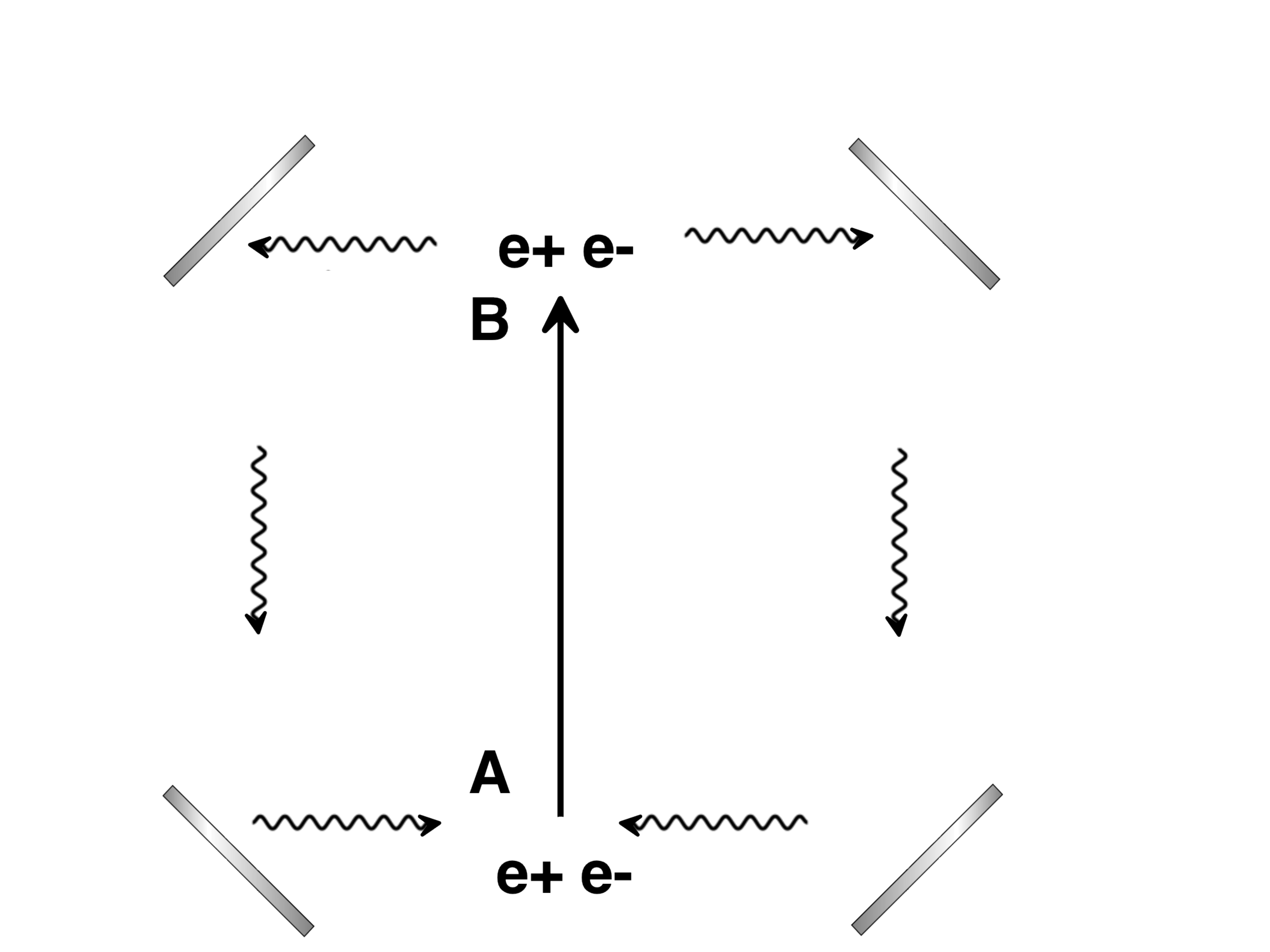}}
\caption{The particle-antiparticle loop of the Morrison argument. The pair starts at point A and then is risen vertically through the gravitational field to point B, where the pair annihilates into two photons. These two photons are propagated down (during which they gain energy) to point A. There they reconvert into a pair, but with an extra photon of energy $2 m_e g \Delta z$, where $\Delta z$ is the vertical distance between A and B.
}
\end{figure}
So the answer to Morrison's argument might well be that the instability associated to antigravity is acceptable since it occurs in most situations at an unnoticeably feeble rate, and is observed in other similar situations such as black hole evaporation. Note, for example, that for a black hole of the mass of our Sun, the evaporation timescale is of the order of $2 \times 10^{66}$ years. As we will see, it is mostly in strong fields, such as those occurring near the horizon of black holes, that this antigravity will lead to significant effects. We will come back to this point as such a vacuum polarization is a prediction of general relativity as soon as we allow the existence of negative mass objects in the vacuum.

Note that some authors have argued, notably in supersymmetric theories, that antigravity, meaning here a slightly different acceleration of antimatter with respect to matter in a gravitational field, is possible without implying any dissipation or instability. This is, in particular, what Jo\"el Scherk proposed in 1979 in another celebrated paper \cite{Scherk} adequately titled ``Antigravity, a crazy idea?"
Scherk argued that in this case, if we assume the existence of gravivector components in $ N = 2, 8$ supersymmetry, we might expect antimatter to fall slightly more rapidly than matter. Still, in the following, we will assume that Morrison's argument was correct, and that the instability is characterized by Hawking's evaporation temperature and therefore acceptable.

\subsection{Schiff's argument}
In his lectures on gravitation \cite{Bell_lectures}, with his second lecture dealing with antigravity, John Bell estimated that probably the most stringent impossibility argument against antigravity was the argument brought by Leonard Schiff at the beginning of the 1960s. The Schiff argument \cite{Schiff} can be summarized in the following way: if antimatter antigravitates, then, depending on the composition of the body used to test the equivalence principle, for example beryllium and uranium, two elements with rather different binding energies, these bodies should follow different trajectories in gravitational fields. To quantify his statement, Schiff tried to estimate the contribution of the virtual electron-positron pairs, and even more importantly the contribution of the virtual quark-antiquark pairs, arguing that pure antigravity would induce variations of the order of these contributions in the acceleration of bodies with different binding energies.

Nieto and Goldman had already noted \cite{Nieto_Goldman} that Schiff's calculation was incorrect as his calculation did not take correctly into account the infinities arising in the renormalisation procedure. Still, we know that, to an outstanding precision, all material bodies, independently of their composition, follow the same trajectories --at least for matter-- for given initial conditions in a gravitational field. The most precise tests have been provided by the experiments of the E\"otwash group \cite{Eotwash} and more recently by the Microscope satellite, the latter providing the most stringent constraint at the level of $2 \times 10^{-14}$ in its first analysis \cite{Microscope}. Even before the Microscope result, with the Schiff argument in mind, it has been estimated (see e.g. \cite{Alves}) that antigravity is constrained at the level of one part per billion. Similarly, Ulmer et al. have stated \cite{Ulmer} that their experiments on the comparison between the proton and antiproton charge over (inertial) mass ratio place a constraint on any gravitational anomaly at a $< 8.7 \times 10^{-7}$ level. Clearly, if this line of reasoning is correct, it is basically useless to perform the AEgIS, ALPHA-g and Gbar experiments, since these experiments are unable to reach the precision required to exceed these constraints.

\subsubsection{Price's argument: General relativity violating the equivalence principle}
Let us see why these statements based on Schiff's argument are probably incorrect: about 25 years ago, Richard Price \cite{Price} studied the behaviour of bound systems composed of a positive and a negative mass in Bondi's sense \cite{Bondi}. He noticed an extremely surprising property: whereas a negative mass falls exactly like a positive mass when it is without interactions, the bound system composed of a mass $+m$ and a mass $-m$, equal and opposite, {\it levitates and polarizes itself} (Fig.~2), the negative mass lying slightly above the positive mass in the levitating system.
\begin{figure}\sidecaption
\resizebox{0.9\hsize}{!}{\includegraphics*{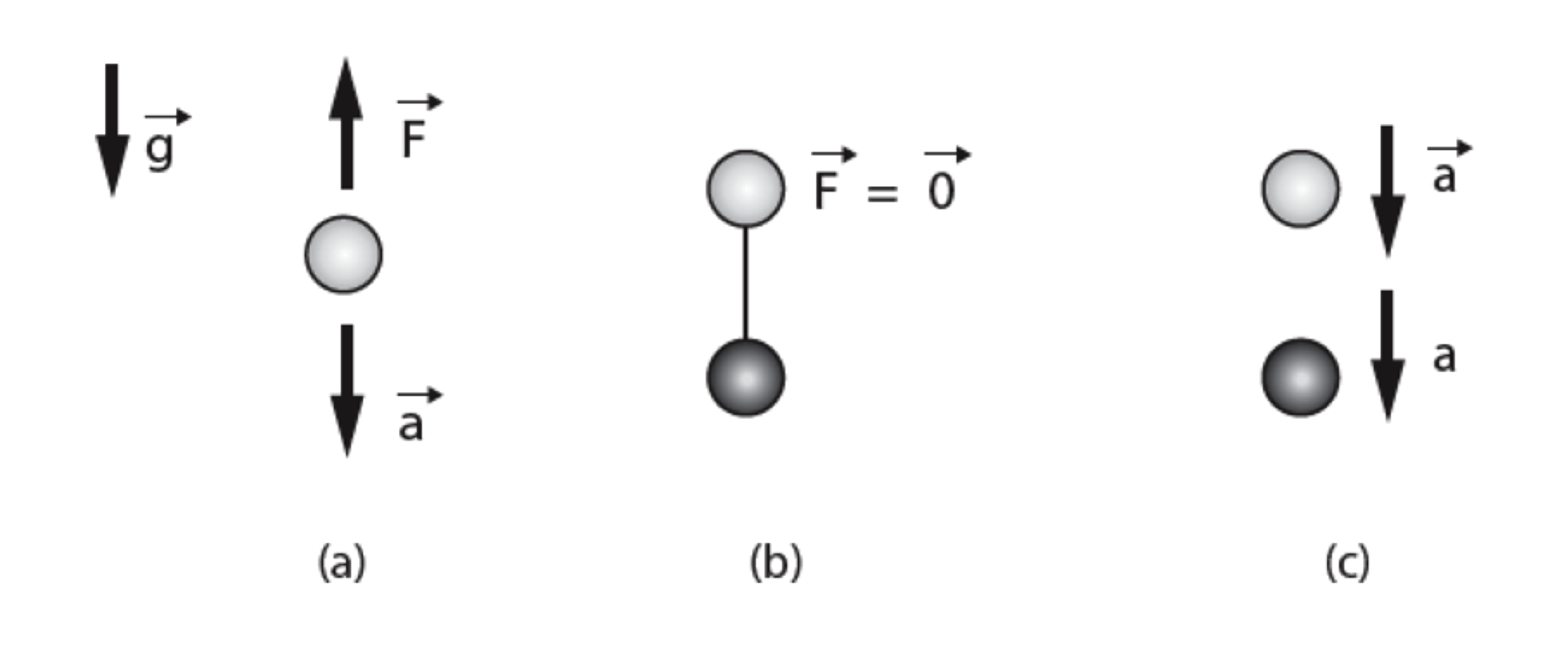}}
\caption{(a) In the Earth gravitational field, although the gravitational force on a negative Bondi mass  $-m$ (where $m$ is positive) is directed upwards, the mass accelerates downwards since its inertial (and gravitational) mass is negative
(b) For a bound system of a $+m$ and $-m$ mass (linked for example by an electromagnetic force), general relativity predicts that the bound system will {\it levitate}
(c) But assuming that the interaction between the two masses can be switched off, both mass fall again at a common pace !
}
\end{figure}

But as soon as the negative and positive masses differ even slightly in absolute value, the composite system always falls with {\it exactly the same acceleration}, respecting again the principle of equivalence: although the overall (inertial) mass of the system has decreased ---it is equal to the algebraic sum of the two masses of the bound system--- and the trajectory of the composite system remains the same.

So, in complete contradiction with intuition, there are only two possible behaviors: a 100\% violation of the principle of equivalence, or no violation at all! This rather clearly, if surprisingly, invalidates Schiff's argument that in case of antigravity, bodies should, according to their composition, undergo (slightly) different accelerations in a gravitational field.

Price also demonstrates that, as soon as negative masses exist, the vacuum will become polarized. While this polarization will go completely unnoticed in weak gravitational fields, such as that of the Earth, in the strong field regime, for example near the horizon of a black hole, the behaviour predicted by general relativity will change dramatically. Indeed, under the diverging tension, the vacuum will eventually break down, and particle-antiparticle pairs will be created. A similar discussion about the breakdown of the vacuum at the horizon of a black hole has occurred in the so-called ``firewall paradox" \cite{Polchinski}, which could find here a solution.

Surprisingly, even at low fields, such a gravitational polarization will lead to observational consequences. Noting that the MOND (Modified Newtonian Dynamics) phenomenological law \cite{Milgrom} had a similar form as the modifications induced by polarization in Maxwell's equations in a dielectric medium, Blanchet and Le Tiec \cite{Blanchet_Le_Tiec} demonstrated that MOND phenomenology could be explained assuming that a gravitational polarization exists. Although they considered initially that this would require a violation of the equivalence principle, we have seen that such a polarization is predicted by General Relativity as soon as negative mass components exist in the vacuum. This would provide a fundamental explanation, within General Relativity, of MOND phenomenology, obviating the need for Dark Matter, definitely missing experimentally but required as a major component in the standard cosmological model.

\subsection{Good's argument}
In 1961, Myron Good \cite{Good} used the neutral kaon system to constrain the different behavior of antimatter with respect to matter in a gravitational field. According to Good, the non-observation of anomalous regeneration (leading to decay in three pions instead of the predominant two-pion decay) in the neutral kaon system imposed very strong constraints on any antigravity, at the $10^{-10}$ level. Good observed that
antigravity would impose that the K$_L$, a linear combination of K$^0$ and its antiparticle, would regenerate a K$_S$ component.
Good estimated the phase shift that would develop between the K$^0$ and its antiparticle from the energy difference due to the gravitational potential
$V_K$; more precisely, he supposed that the phase factor between the two
components would oscillate as:
$$\exp (i m_K V_K t/\hbar)$$
where $m_K$ is the mass of the kaon and $V_K$ is the potential energy of the kaon.
Noticing that the potential energy of a kaon in the
Earth gravitational field is $\approx 0.4$ eV, and that this energy
is $\approx 10^5$ times larger than the energy splitting between
the K$_S$ and K$_L$ eigenstates, Good concluded that antigravity
was constrained by the non observation of three-pion decay at the $\approx 10^{-10}$ level.

This argument suffers, however,
from a severe criticism: as Good himself had
noticed, there is no obvious reason why one should
use the Earth potential; why not use instead the Sun,
or the galactic potential which would give even more
stringent limits on the difference of acceleration between
matter and antimatter? In fact, as noted by Nieto and Goldman \cite{Nieto_Goldman}, an e$^+$--e$^-$ pair created in a deep potential well in the so-called Klein paradox \cite{Holstein} shows that the phase difference between the electron and the positron builds up with their separation instead of being created instantaneously. Supposing that, when at rest with respect to one another, the neutral kaon and its antiparticle have different frequencies in their phase factors is equivalent to saying that the $K^0$ and its antiparticle do not have the same mass -- i.e., a most severe CPT violation.
Chardin and Rax \cite{Chardin_Rax} and Goldman et al. \cite{Goldman_Nieto_Sandberg} restated Good's argument independently
of absolute potentials, assuming that a particle and its antiparticle have the same frequencies (and the same mass) when they are at rest with respect to one another. They found just the opposite conclusion: antigravity predicts the approximate amount of anomalous regeneration associated with CP violation, discovered three years after Good had proposed his argument \cite{cpviolation}. Note that Bell and Perring \cite{Bell_Perring}, immediately after the discovery of CP violation, had reversed the Good argument to invoke a cosmological field differentiating matter and antimatter. 

\section{The no-go theorem for symmetric matter-antimatter cosmologies}
\label{sec:nogo}

The Dirac-Milne Universe that we will discuss in Sec. \ref{sec:diracmilne} is a universe that
contains as much matter as antimatter.
In the 1960's and again in the late 1980s, two teams tried
to understand whether a symmetric matter-antimatter universe could be consistent with observations.
During the 1960's, the group led by
Roland Omn\`es \cite{Omnes}, at University of Paris-Sud, made the assumption that at the time of the quark-gluon plasma transition, at a temperature of
about 170 MeV, a matter-antimatter emulsion was formed, which developed through annihilation at the matter-antimatter boundaries.
The conclusion of their study, after several years of effort, was that the primordial Universe, at least in
the standard cosmological model, does not provide enough time for a matter-antimatter universe to create structures large enough to evade the constraints of the diffuse gamma-ray flux. Indeed, if
matter and antimatter have both positive gravitational masses
and respect the equivalence principle,
matter-antimatter annihilation continues to occur
after the Universe has become transparent, leading to a
diffuse high energy gamma-ray background, already difficult
to justify in the early 1970s, at the epoch of the SAS-2 satellite, and clearly inconsistent
with the much more sensitive contemporary satellites, such as the Fermi satellite.

In the late 1980s and early 1990s, Sheldon Glashow, Andrew Cohen and Alvaro de
Rujula \cite{Cohen_et_al} took over the study of Roland Omn\`es' group,
focusing on the late periods when the universe has become transparent,
about 380,000 years after the Big Bang
in the standard cosmological model.
They concluded again that even if the primordial universe
had succeeded in creating a
symmetrical world between matter and antimatter, diffusion
of matter and antimatter at their domain boundaries
would lead, as soon as the Universe becomes
transparent, to an annihilation between matter and
antimatter in conflict with observational limits, unless the domains of
matter and antimatter had a size exceeding several billion light-years,
which seems unrealistic.

But rather obviously,
these studies had assumed that matter and antimatter had
both a positive gravitational mass. Today, the
discovery of a repulsive gravity through
observations of SN1a supernovae luminosity distance, called Dark Energy,
leads us to be more cautious: indeed, what seemed
previously impossible, repulsive gravity, is the major topic of interest in cosmology,
and we now turn to the study of negative mass and its definition.

%% For one-column wide figures use
%\begin{figure}
%% Use the relevant command to insert your figure file.
%% For example, with the graphicx package use
%  \includegraphics{example.eps}
%\end{figure}
%%
%% For two-column wide figures use
%\begin{figure*}
%% Use the relevant command to insert your figure file.
%% For example, with the graphicx package use
%  \includegraphics[width=0.75\textwidth]{example.eps}
%\end{figure*}
%\begin{table}
%\begin{tabular}{lll}
%\hline\noalign{\smallskip}
%\end{tabular}
%\end{table}

\section{Negative mass}
\label{sec:negmass}
Before presenting the main features of the Dirac-Milne matter-antimatter universe that was studied from 2006 on by Benoit-L\'evy and Chardin  \cite{Benoit-Levy_Chardin}, it is useful to go back to the meaning that can be given to the notion of negative mass.

By the 1950s, Bondi \cite{Bondi} had built negative mass solutions that respected the equivalence principle. The surprising properties of these solutions, for example the ``runaway" motion when two equal but opposite masses accelerate continuously while remaining at (almost) constant distance, led them to be considered very exotic objects, although such runaway motions can also be observed in situations involving only positive masses. In addition, very strong theorems seemed to exclude any possibility of negative mass particles or objects, or more generally violating the positivity of energy \cite{Schoen_Yau,Witten}.
But whereas initially these theorems on the positivity of energy appeared as absolute no-go theorems, the increasing number of violations of the energy conditions, first through the quantum effects of the vacuum (for example in the Casimir effect), then from 1998 in an infinitely more significant way with the discovery of Dark Energy \cite{Perlmutter_1999,Riess_1998}, led to question these theorems. For a review on counter-examples of the various expressions of energy conditions, see for example the review by Barcelo and Visser \cite{Barcelo_Visser}.

The demonstration of the existence, in 2014, of perfectly respectable solutions of negative mass ``bubbles" without instability \cite{Paranjape} as soon as they are placed in an expanding universe (here, the Einstein-de Sitter universe) finally demonstrated that instability does not constitute a sufficient argument to exclude a solution: it is indeed also necessary to calculate  the characteristic time of instability, since cosmological solutions are themselves unstable, but with often enormous and therefore acceptable characteristic times.

In other words, if a negative energy solution is unstable in Minkowski (flat) spacetime, but is stable in an Einstein-de Sitter spacetime, while a few billion years are required to determine whether you live in one or the other of these two universes, it means that the instability of the negative mass solution has at most a characteristic timescale of a few billion years.

Also, the analysis of Klein's ``paradox" \cite{Holstein} shows that, since fermions always come in pairs, the vacuum breaks down with pair creation when a electron of mass $m_e$ is confined in a potentiel well of depth larger than $-2m_e c^2$ and {\it not} $m_e c^2$, so that the electron has a negative total energy of $-m_e c^2$ when the vacuum starts to break down.

Now, Dirac had shown that antimatter appears as the matter of negative energy going backwards in time. And we also know since the early 1990s that building a time machine in general relativity ---for example using a wormhole as a time machine--- requires violating the positivity of energy \cite{Morris_Thorne_Yurtsever}. It is therefore natural to test whether antimatter is not by any chance such an ``exotic" material.

Interestingly, Hawking had noted \cite{Hawking_Mach} that the Machian formulation of general relativity proposed by Hoyle and Narlikar \cite{Hoyle_Narlikar} only works if there exists equal amounts of negative mass and positive mass particles in the universe. For Hawking, it clearly meant that the theory was wrong, but today, with the knowledge that negative mass solutions are allowed, and the observation of repulsive gravity, it is fascinating to see that the initial Machian perspective of Einstein could effectively be realized in the Dirac-Milne universe, which we will describe in the next section.

\section{Dirac-Milne cosmology}
\label{sec:diracmilne}
The discovery in 1998 of a mysterious repulsive energy, dubbed Dark Energy, and representing more than two thirds of the energy content of the universe, provided the first massive evidence for repulsive gravity. It also underlined the improbability of the Standard Model of Cosmology, featuring an extremely brief initial phase of very brutal deceleration, followed by a rather mysterious and very brief repulsive phase of inflation, mostly justified by the need to solve the enigma of the homogeneity of the primordial universe. In particular, there does not exist any precise fundamental theory allowing to understand how one can not only enter but also leave this phase of inflation. A time slightly too long will lead to a virtually empty universe, whereas a time slightly too short will lead to the re-collapse of the universe in a few Planck times ($10^{-44}$ s).

At the end of inflation, when the universe is not even $10^{-30}$ seconds old, a new phase of very violent deceleration is supposed to start, leading, about $10^5$ years later (virtually an eternity compared to the two previous epochs), to a period where matter, until then mostly irrelevant, manages eventually to become the majority component while radiation as well as dark energy are then totally negligible. And it is only at the age of a few billion years that the Dark Energy component, a name hiding our ignorance of its true nature, becomes dominant, leading to a universe of accelerated expansion, where galaxies will find themselves isolated from each other in a (relatively) near future.

Another major drawback of the Standard Model is the fact that it uses two predominant components, dark matter and dark energy, supposed to represent about 95\% of the universe energy density, but which have remarkably resisted to experimental identification so far.
Quoting the Planck HFI collaboration \cite{Planck_2018}, ``the six-parameter Lambda-CDM model continues to provide an excellent fit to the cosmic microwave background data at high and low redshift, describing the cosmological information (...) with just six parameters (...). Planck measures five of the six parameters to better than 1\% accuracy (simultaneously), with the best-determined parameter ($\theta_*$) now known to 0.03\%."

But an excellent fit to the data at a given epoch is not a guarantee of a correct description of reality. 

Indeed, several authors have noted \cite{Nielsen_et_al,Tutusaus_et_al} that our universe is very similar to a gravitationally empty or coasting universe (neither accelerating or decelerating), which was first envisaged by Milne \cite{Milne}. On this basis, Benoit-L\'evy and Chardin  \cite{Benoit-Levy_Chardin} proposed the so-called ``Dirac-Milne" universe,
a universe containing the same amount of matter and antimatter (hence Dirac's name), endowed respectively with positive and negative mass. Like Milne's, this is a cosmology that is permanently on the verge of inflation and therefore able to explain the initial homogeneity of the Universe.

Although a Milne universe -- advocated for example by Melia in his $R_H = ct$ universe \cite{Melia}, without antimatter -- suffers from the depletion of deuterium and helium-3 abundance and a widely different CMB angular scale, the Dirac-Milne universe is impressively concordant: in addition to the age of the Milne universe, equal to $1/H_0$,  almost exactly that of the Lambda-CDM universe, and a SN1a luminosity distance (Fig. 3) also impressively similar, the CMB angular scale originating from the sound of the matter-antimatter annihilation is at the one-degree scale, and primordial nucleosynthesis is reproduced, including deuterium.

\begin{figure}\sidecaption
\resizebox{0.9\hsize}{!}{\includegraphics*{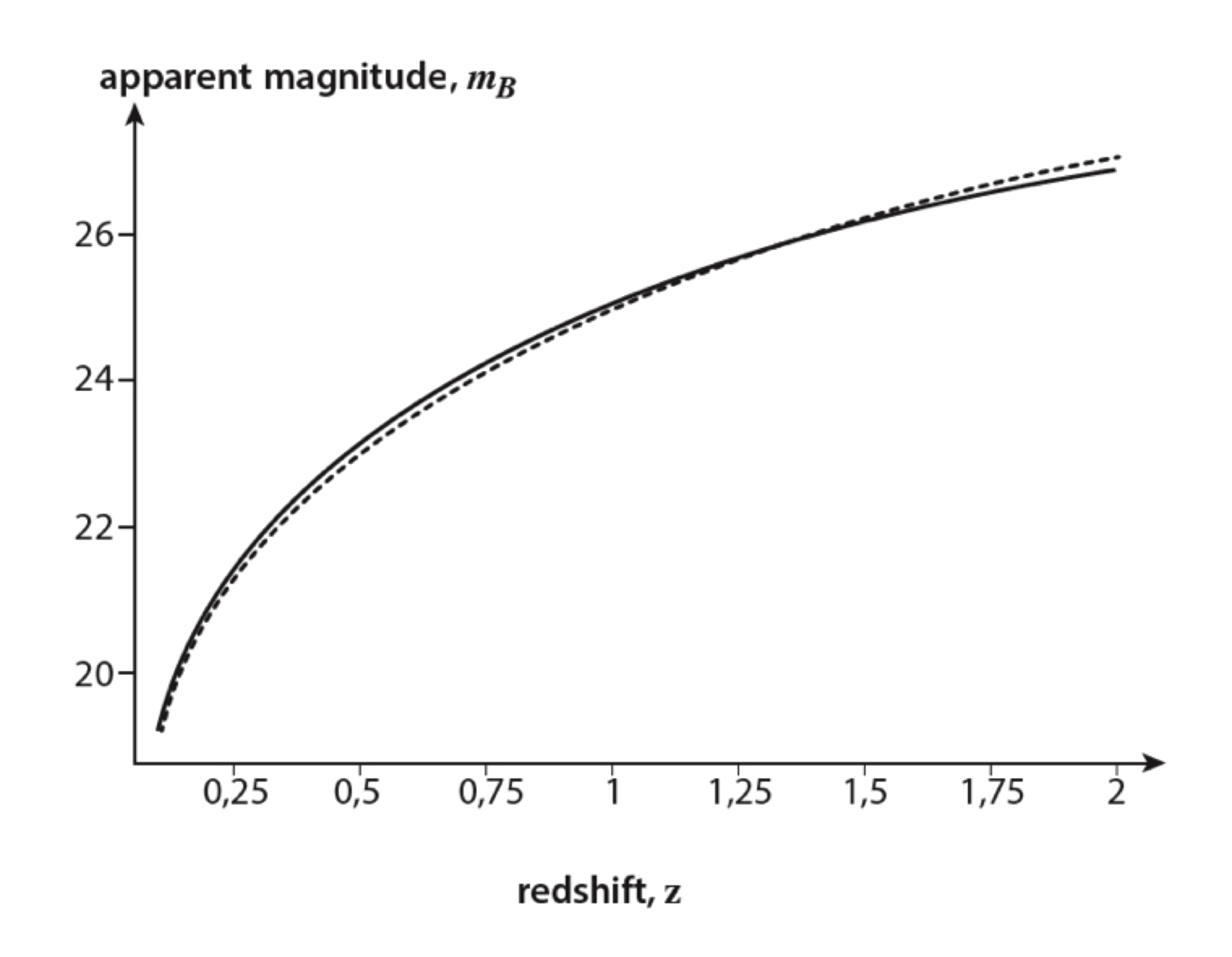}}
\caption{Magnitude-redshift diagram for SN1a supernovae. The solid line shows the prediction of the Lambda-CDM cosmology with $\Omega_m = 0.28$ and $\Omega_{\Lambda} = 0.72$, while the dotted line represents the prediction of the Dirac-Milne cosmology. Clearly, it will be extremely difficult to distinguish the two cosmologies using SN1a supernovae. Adapted from Chodorowski \cite{Chodorowski}.
}
\end{figure}

Taken at face value, there are two observational problems facing the Dirac-Milne universe. The first problem is related to an overproduction, by nearly a factor 10, of helium-3. But helium-3 is a bad probe of the primordial universe since it can be both destroyed and produced after the initial phases of the universe. As a result, it has sparked little enthusiasm from experimentalists, and the measurements, clearly below the prediction of the Dirac-Milne universe, are probably inconclusive since the authors themselves note \cite{Science_Helium3} that (their) ``result for $^3$He is exactly the {\it opposite} of what one would expect (...) The utility of $^3$He/H as a probe of the cosmological baryon-to-photon ratio rests on the resolution of this puzzle."

The second problem, that we address in the following section, is provided by the Baryonic Acoustic Oscillations (BAO), observed at the present comoving scale of $\approx 100$ Mpc, and which have no direct equivalent in the Dirac-Milne cosmology. We now proceed to show that such a scale is produced without any free parameter in the non-linear regime of structure formation in the Dirac-Milne universe.

\subsection{Structure formation in the Dirac-Milne cosmology}
Motivated by the fact that our universe shares several aspects with a coasting universe, we have studied structure formation in universes involving equal amounts of negative and positive mass, with a particular emphasis on the Dirac-Milne cosmology \cite{Benoit-Levy_Chardin}. The first results of these simulations have been presented by Manfredi at this conference, and since then published in a more detailed paper \cite{Manfredi_et_al_PRD}. We summarize here briefly the results of these first simulations, and the reader is referred to the publication \cite{Manfredi_et_al_PRD} for more detailed information about structure formation involving negative mass.

Dubinski and Piran \cite{Dubinski_Piran}, and later Piran \cite{Piran} had noted in the early 1990s that the evolution of underdense regions, which expand and lead to large ``voids" that occupy the largest fraction of our universe, could be described as the evolution of negative mass particles violating ``maximally" the equivalence principle. As noted above, this violation of the equivalence principle has a physical motivation as it corresponds to the electron-hole system in a semiconductor.

Surprisingly, the Dirac-Milne scenario cannot be recovered by simply assigning a combination of signs to the three types of Newtonian masses, i.e., the inertial, active gravitational and passive gravitational masses. Instead, one needs to resort to a bimetric formalism, which in the Newtonian limit reduces to a set of two Poisson's equations for the gravitational potential.

With this bimetric formalism, starting from the initial conditions dictated by the Dirac-Milne universe and the evolution of the matter-antimatter emulsion during the initial stages of the universe, at temperatures higher than $\approx 30$ eV, our Newtonian simulations show the gradual buildup of structures. 
Such structures begin to develop a few million years after the CMB transition, much earlier than in the Lambda-CDM standard model, then grow in size, reaching a maximum comoving size of $\approx 100$ Mpc (Fig. 4) a few billion years after the Big Bang.
This size is characteristic of the BAO scale, and could provide an explanation for the otherwise unexpected coincidence between the linear BAO fixed comoving scale ---supposed to provide a standard ruler--- and the evolving non-linear scale, observed for example in SDSS \cite{SDSS}. This provides a new element of concordance between the Dirac-Milne universe and our universe, and a further motivation to pursue a more detailed study of this matter-antimatter universe.

\begin{figure}\sidecaption
\resizebox{0.9\hsize}{!}{\includegraphics*{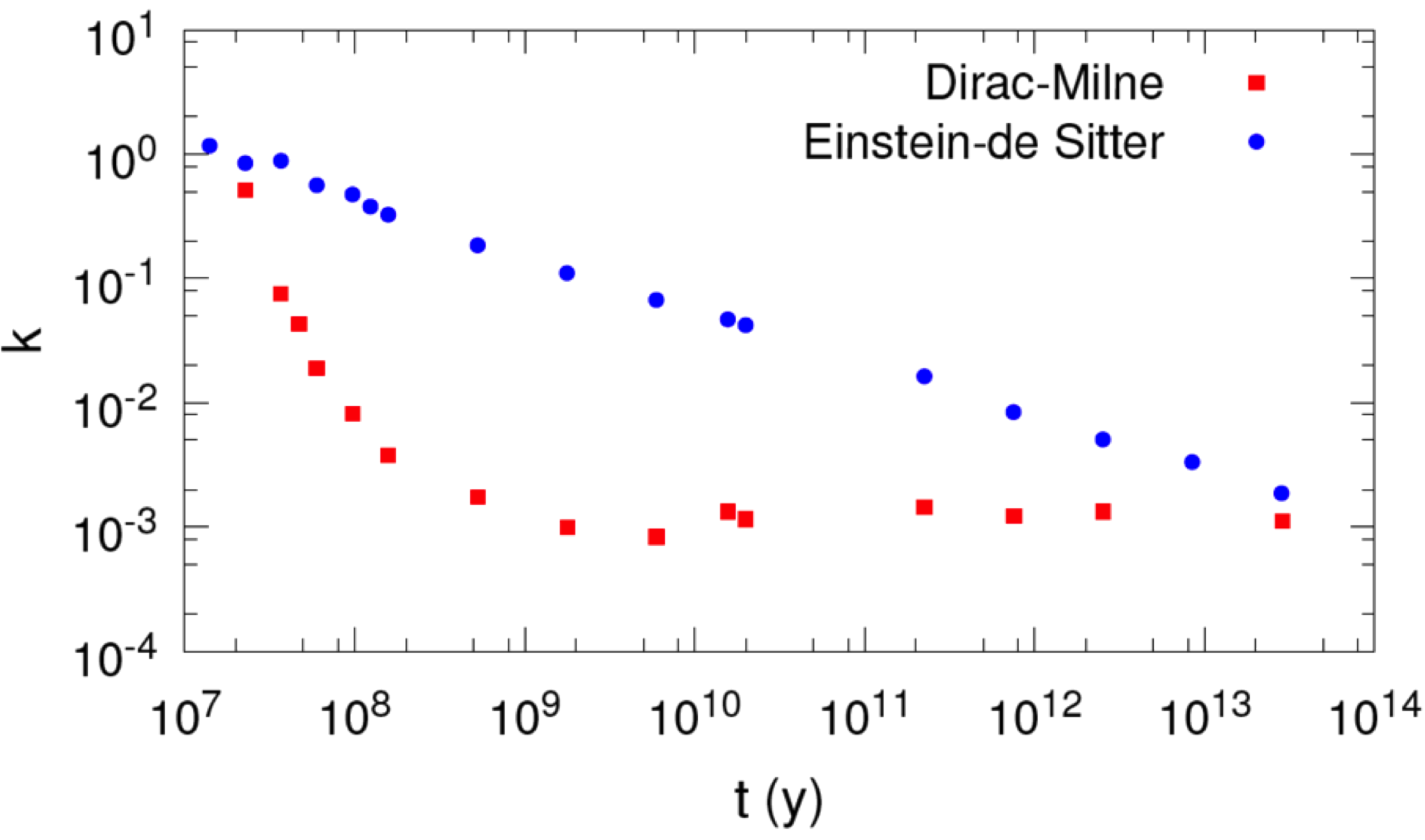}}
\caption{Wavenumber corresponding to the peak in the power spectrum for the Dirac-Milne and Einstein-de
Sitter universes as a function of time, in comoving coordinates.
}
\end{figure}

\subsection{Other tests of the Dirac-Milne cosmology}
Further lines of study can be realized to test the validity of the Dirac-Milne cosmology. As a hypothesis that remains to be confirmed, the Dirac-Milne cosmology may also explain why there remains approximately only one billionth of matter following the primordial annihilation, with the same amount of antimatter surviving in clouds of cold gas occupying the vast majority of the intergalactic space in our universe.

The calculation of the matter-antimatter annihilation occurring between the temperature $T=170 \,\rm MeV$, corresponding to the quark-gluon plasma transition and $T=30 \,\rm eV$, where the matter regions separate from the antimatter regions, could provide a key to understand and calculate the parameter $\eta = n_{baryon}/n_{photon}$, for which various mechanisms have been proposed, but without providing any prediction of its precise value, which is essential to our very existence.
Similarly, it would also seem interesting to further study the possible explanation proposed by Blanchet and Le Tiec \cite{Blanchet_Le_Tiec} that gravitational polarization might provide an explanation for MOND, mimicking the existence of the evasive Dark Matter.

\section{Conclusions}
\label{sec:conclusions}
Although most physicists would still bet against antigravity for antimatter, the situation has changed rather dramatically since the discovery in 1998 of what we call, by lack of better understanding, Dark Energy, representing about two thirds of the universe energy density.
 Considering that another ``dark" component, namely Dark Matter, is supposed to represent roughly 25\% of the universe, the standard cosmological model finds itself in the unpalatable situation of explaining most observations using concepts that are little, or not at all, understood. It is therefore reasonable to investigate possible alternatives to the standard model, which may in the end turn up to be just an impressive fit to the data using a relatively limited number of parameters.

The fact that in several respects our universe appears very similar to a coasting or empty universe is a motivation to reconsider the impossibility arguments against the existence of negative mass, on the one hand, and antigravity, on the other. This led us to the study of the properties of the Dirac-Milne universe, a symmetric matter-antimatter universe, where antimatter is endowed with negative mass, an analog of the electron-hole system in a semiconductor, providing a cosmology impressively concordant with our universe (age, SN1a luminosity distance, nucleosynthesis, structure formation, CMB). Although much remains to be done on the CMB front, where the full sound spectrum should be established, the present results are a strong motivation to deepen the study of this cosmology.

Importantly, a key test of the Dirac-Milne cosmology will soon be realized in the laboratory with cold antihydrogen atoms: the AEgIS, ALPHA-g and Gbar experiments at CERN are expected to provide tests of the antigravity hypothesis. While Gbar is preparing a precision measurement with antihydrogen ions cooled to a few tens of microkelvin, the ALPHA collaboration, which made the first spectroscopic measurements on antihydrogen \cite{ALPHA_spectroscopy}, achieved in 2013 the first constraints on antimatter gravity \cite{ALPHA_gravity}, though still a factor $\approx 65$ larger than the sensitivity necessary to test antigravity.

Whatever the experimental results, they will have important repercussions on our understanding of the evolution of our universe.

%\begin{acknowledgements}
%\end{acknowledgements}



\end{document}